\documentclass[letterpaper,twocolumn,10pt]{article}
\usepackage{usenix,epsfig,endnotes}
\usepackage{color}
\usepackage{graphicx}
\usepackage{amssymb,amsmath}
\usepackage{subfigure}
\usepackage{balance}
\usepackage{cite}
\usepackage{url,xspace}
\usepackage{mdwlist}
\usepackage{paralist}
\usepackage{multirow}
\usepackage{flushend}
\begin{document}

%don't want date printed
\date{}
\graphicspath{{./figures/}}
%make title bold and 14 pt font (Latex default is non-bold, 16 pt)
\title{\Large \bf Thermal Covert Channels on Multi-core Platforms}

%for single author (just remove % characters)
\author{
{\rm Ramya Jayaram Masti$^*$, Devendra Rai$^\dagger$, Aanjhan Ranganathan$^*$, Christian M{\"u}ller$^\dagger$} \\ {\rm Lothar Thiele$^\dagger$, Srdjan Capkun$^*$}\\
$^*$Institute of Information Security, ETH Zurich \\
$^\dagger$Computer Engineering and Networks Laboratory, ETH Zurich 
%\and
%{\rm  Devendra Rai, Christian M{\"u}ller, Lothar Thiele}\\
%Computer Engineering and Networks Laboratory, ETH Zurich 
% copy the following lines to add more authors
% \and
% {\rm Name}\\
%Name Institution
} % end author

\maketitle

% Use the following at camera-ready time to suppress page numbers.
% Comment it out when you first submit the paper for review.
\thispagestyle{empty}

\newcommand{\todo}[1]{\textcolor{red}{TODO: #1}}
\newcommand{\citneed}{[\textcolor{red}{cit}] }
\newcommand{\unit}[1]{\ensuremath{\, \mathrm{#1}}}
\newcommand{\eg}{e.g.,\xspace}
\newcommand{\bigeg}{E.g.,\xspace}
\newcommand{\etal}{\textit{et~al.\xspace}}
\newcommand{\etc}{etc.\@\xspace}
\newcommand{\fixit}[1]{\textcolor{red}{#1}}
\newcommand{\ie}{i.e.,\xspace}
\newcommand{\bigie}{I.e.,\xspace}
\newcommand{\bitlen}{T$_{b}$}
\newcommand{\timeslice}{t$_s$}

\subsection*{Abstract}
Side channels remain a challenge to information flow control and security in
modern computing platforms. Resource partitioning techniques that minimise the
number of shared resources among processes are often used to address this
challenge. In this work, we focus on multi-core platforms and we demonstrate
that even seemingly strong isolation techniques based on dedicated cores and
memory can be circumvented through the use of thermal side channels.
Specifically, we show that the processor core temperature can be used both as a
side channel as well as a covert communication channel even when
the system implements strong spatial and temporal partitioning. Our experiments
on an x86-based platform demonstrate covert thermal channels that achieve up
to 12.5 bps and a weak side channel that can detect processes executed on neighbouring
cores. This work therefore shows a limitation in the isolation that can be
achieved on existing multi-core systems.

\section{Introduction}
Side channels have for a long time remained an open threat to information flow control and isolation
techniques in a variety of contexts including cloud and mobile
computing~\cite{wu-usenix12, xu-ccsw11,zhang-ccs14, marforio-acsac12}. Such channels can be used for data exfiltration
from a victim~\cite{zhang-ccs14} and  be exploited by colluding applications
to covertly exchange information~\cite{wu-usenix12,ristenpart2009hey}.

A common technique to mitigate side channel attacks that leverage co-location is
to dedicate resources (e.g.,~processor cores, memory) to individual processes for the duration of their
execution.  Although seemingly inefficient, such a technique is becoming viable
with the appearance of multi and many-core systems which contain hundreds of cores~\cite{xeonphi, tilera, adapteva}. However, it has already been shown that, due to their
architecture and use, multi-core systems do not trivially protect against all types of information leakage --- side channels have been demonstrated leveraging shared caches~\cite{wang-acsac06}, memory bus~\cite{wu-usenix12}, network stacks~\cite{mileva-jcs14}, virtual memory~\cite{suzaki-eurosys11}, I/O devices~\cite{shah-usenix06} etc. These
side-channels, however, still exploit the resources that particular multi-core
platforms, for performance and other reasons, share among the
processes. This resulted in proposals that solve this problem by
partitioning the shared resources when possible, for example, dividing caches~\cite{wang-acsac06} and
bus bandwidth~\cite{gundu-hasp14}.  

In this paper, we show that even strong isolation techniques based on dedicated cores and memory can be circumvented in multi-core systems through the use of a \emph{thermal
  side-channel}. For this we leverage the thermal properties of multi-core systems and access to core temperature information that, for performance reasons, is available to processes on these platforms. First, the thermal capacitance and resistance of computing platforms result in remnant heat from computations, i.e., the heat is observable even after that computation has stopped. As a result, information about one process may leak to another that follows it in the execution schedule. Second, the effects of heat resulting from
processes running on one core can be observed on other cores across the chip.
This leaks information about a process to its peers running on other cores in a
processor chip. We demonstrate our attacks on commodity multi-core systems.  So far, thermal
(heat) side channels have not been studied on these systems. There is a trend
towards exposing such data to users and allowing them to make thermal management decisions based on it~\cite{brown-linux07}. For example, temperature information is
accessible from user-space on modern Linux systems~\cite{coretemp}. This paper highlights the tension
between building thermally-efficient systems which requires exposing high-quality
temperature data to applications and securing them. 

In summary, we make the following contributions: (i) We demonstrate the feasibility of using thermal side channels for communication between colluding applications and measure the throughput of such a channel on an Intel-based platform. The challenges in building such a channel include the system's thermal capacitance, effect of cooling on multi-core systems and resolution limitations of the thermal sensors available on these platforms. (ii) We explore the factors that influence the throughput of this side channel: frequency and relative locations of the colluding applications (processes). (iii) We demonstrate the existence of limited thermal side channel leakage from processes running on adjacent cores that allow identification of applications based on their thermal traces. On existing systems, heat-based leakage is non-trivial to avoid without a performance penalty; we discuss possible countermeasures to eliminate or limit the impact of such attacks.

The rest of this paper is organised as follows. In Section~\ref{sec:background}, we discuss the background and motivation for our work. Section~\ref{sec:therm-behavior} discusses the thermal behaviour of x86 platforms and Section~\ref{sec:intro-channels} describes how these properties can be exploited to create thermal channels. In Section~\ref{sec:covert}, we demonstrate the feasibility of using thermal side channels for covert communication even in systems with isolation based on spatial and temporal partitioning. Section~\ref{sec:profiling} demonstrates that limited side channel leakage can occur through thermal channels which can be exploited for unauthorised application profiling. Section~\ref{sec:discussion} and Section~\ref{sec:related-work} summarise countermeasures against thermal channels and related work respectively. Finally, we conclude in Section~\ref{sec:conclusion}.

\section{Background and Motivation}
\label{sec:background}
In this section, we summarise the use of thermal information in modern processors and resource partitioning-based isolation techniques, as well as give a motivation for our study.

\subsection{Thermal Management}
Thermal management is key to the safe and reliable operation of modern computing systems. Today, thermal sensors are incorporated into a number of system components including hard-drives, DRAM, GPU, motherboards and the processor chip itself~\cite{tempsensors}. In this work, we focus on the information available from thermal sensors that are embedded in processor chips.

Ensuring the thermal stability of a processor is becoming increasingly challenging given the rising power-density in modern processor chips. As a result, major processor vendors (e.g., Intel, AMD, VIA) incorporate thermal sensors to enable real-time monitoring of processor temperature. ARM-based processors also include thermal sensors inside the system-on-chip (SoC) for power and temperature management.

Initially, thermal management was done statically in hardware and included mechanisms to power-off the processor to prevent melt-downs. This later evolved to more sophisticated dynamic frequency and voltage scaling techniques that change processor frequency to lower its temperature~\cite{amd-cool,intel-speedstep}. Hybrid software- and hardware- approaches to thermal monitoring have become popular over time; operating systems today poll temperature sensors and use this to manage cooling mechanisms such as processor frequency scaling and fan-speed~\cite{linux-governors}. More recently, there is a trend towards user-centric thermal management that exposes thermal data to users and allows them to implement customised thermal management policies. For example, Linux-based systems today enable users to configure thermal policies~\cite{brown-linux07, linaro}.

The number and topology of thermal sensors depend on the processor vendor and family. For example, while Intel and VIA processors expose temperature data for individual cores using on-die sensors,  some AMD (e.g., Opteron K10 series) processors only allow monitoring the overall temperature of the entire chip using a sensor in the processor socket~\cite{coretemp}. Optimising the number and placement of thermal sensors on processors is an active research topic~\cite{nowroz-dac10,mukherjee2006systematic,lee2005analytical,memik2008optimizing}.

\subsection{Resource Partitioning-based Isolation}
Isolation techniques for multi-core platforms that are based on resource partitioning offer a number of benefits.  First, resource management approaches that rely on partitioning reduce the size of the software Trusted Computing Base (TCB)~\cite{nohype-ccs11}. In fact, some modern servers (e.g., from Hitachi) incorporate such hardware resource partitioning functions into firmware~\cite{virtage} and create multiple, independent execution containers that can co-exist securely without the need for a software TCB. Second, the simplicity of partitioning-based resource management eases formal verification and this is leveraged by separation kernels like Muen~\cite{buerki-techreport13}. Third, modern processors rely on partitioning techniques to build Trusted Execution Environments (TEEs). TEE technologies such as Intel Trusted Execution Technology (TXT)~\cite{txt}, Intel Software Guard Extensions (SGX)~\cite{sgx1} and ARM TrustZone~\cite{trustzone} create isolated containers for the execution of sensitive software. Intel TXT relies on temporal partitioning of resources such as CPU and memory between trusted and untrusted software. Intel SGX and ARM TrustZone use temporal partitioning only for the CPU and implement spatial partitioning for memory resources in the system.

Resource partitioning has already been proposed as a countermeasure against
side-channels\cite{gligor-dtic93}. This is based on the understanding that
processes can modify shared resources (e.g., cache, file) to communicate
covertly or exfiltrate sensitive information from victim processes by tracking the state of a shared resource such as cache. Therefore, there have been proposals for minimising shared resources to reduce the number and
effectiveness of side channels. Examples include partitioning of caches~\cite{wang-acsac06} and
and bus-bandwidth~\cite{gundu-hasp14}. 

\subsection{Motivation}

The main motivation behind this work is our observation that despite its security
advantages, resource partitioning on multi-core systems might not be able to completely eliminate some types of
inference or communication across partitions. More specifically, we want to investigate if the exposure of core temperature information 
could be used to build both side channels and covert communication channels between processes that execute 
on different cores within a multi-core system. Our goal is to study these channels primarily in terms of their feasibility and throughput. 

Such channels are particularly interesting in the context of multi-core systems for two main reasons: \emph{(i)} today, 
 these platforms expose the information from thermal sensors to users and \emph{(ii)} they can be tested for their effectiveness under the resource partitioning-based isolation mechanisms that multi-core systems can support. To build thermal side channels, it is necessary to understand the type and quality of temperature data available on systems today. One must also account for the nature
of temperature variations on such systems and the factors that affect
them. We focus on Intel x86 platforms for our study given their wide spread use. 

\section{Thermal Behaviour of x86 Platforms}
\label{sec:therm-behavior}
In this section, we first describe the on-die thermal sensors available on Intel processors and then discuss the thermal behaviour of these platforms under a CPU-intensive load. 

\subsection{Temperature Sensors in Intel \\Processors}
Intel labels each of its processors with a maximum \emph{junction temperature} which is the highest temperature that is safe for the processor's operation. If the processor's temperature exceeds this level, permanent silicon damage may occur. To avoid such processor melt-down, Intel facilitates processor temperature monitoring by incorporating one
digital thermal sensor (DTS) into each of the cores in a processor. The layout of the cores within a processor chip can be identified using \texttt{lstopo}~\cite{lstopo}. For example, on the Xeon server used in our experiments, the cores are arranged along a line (as shown in Figure~\ref{fig:setup}). Each DTS reports the
difference between the core's current temperature and the maximum junction
temperature~\cite{dts}. The accuracy of the DTS varies across different generations
of Intel processors. They typically have a resolution of~$\pm$1$^{\circ}$C.

The absolute value of a core's temperature in $^{\circ}$C is computed in software by
subtracting the thermal sensor reading from the maximum junction temperature.
Thermal data from a sensor can be obtained using special CPU registers of the corresponding core.
The data from all sensors is
exposed using the \texttt{coretemp} kernel module~\cite{core temp} on Linux systems and is accessible from user space 
through the \texttt{sysfs} filesystem which is refreshed every 2 ms.

\begin{figure}[t]
  \centering
 \includegraphics[width=0.99\columnwidth]{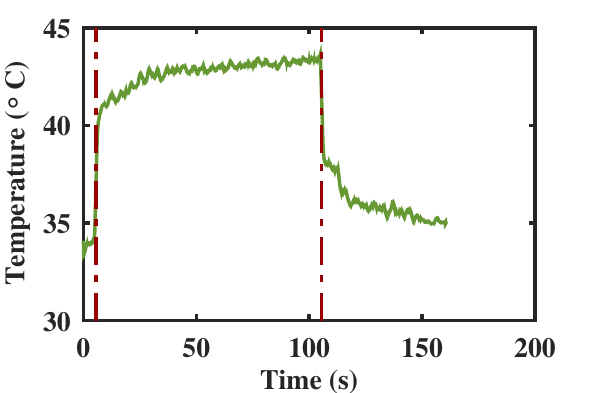}  
 \caption{\textbf{Thermal response of a CPU intensive application:} Temperature
   trace resulting from the execution of an application that does RSA decryption
   in a loop for 100 s. The start and end times of the application's
   execution are indicated using the two red lines. Temperature increases rapidly initially and saturates over time as
   an application runs. Similarly, it falls rapidly as soon as the core becomes idle and gradually returns to the ambient temperature.}
  \label{fig:single-freq}
\end{figure}
\begin{figure*}[t]  
  \centering

 \subfigure[\textbf{1 Hop}]{ \includegraphics[width=0.4\textwidth]{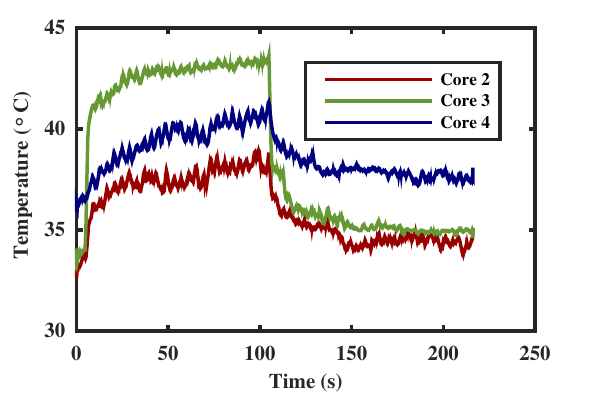}  \label{fig:hop1}}
 \subfigure[\textbf{2 Hops}]{\includegraphics[width=0.4\textwidth]{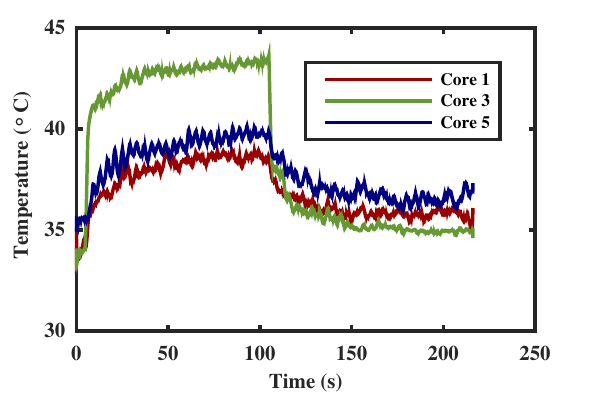}\label{fig:hop2}}
 \subfigure[\textbf{3 Hops}]{ \includegraphics[width=0.4\textwidth]{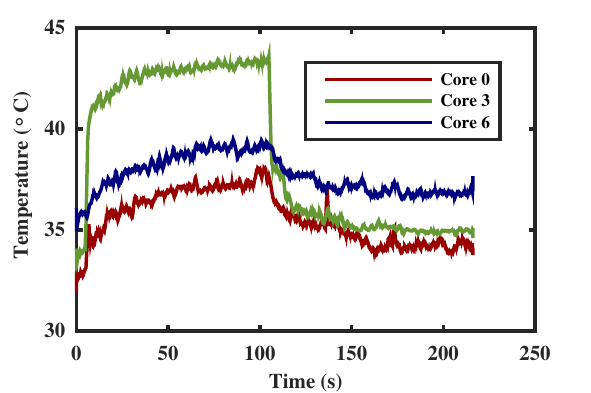}  \label{fig:hop3}}
 \subfigure[\textbf{4
   Hops}]{\includegraphics[width=0.4\textwidth]{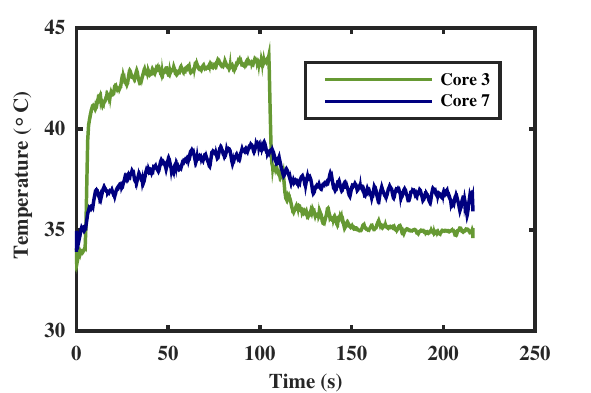} \label{fig:hop4}}  
\caption{\textbf{Heat Propagation from a Neighbouring Core:} Effect of running a
  CPU-intensive application on core~3 of an octa-core processor for 100~s on the temperature sensors of its adjacent cores.}
  \label{fig:hops}
\end{figure*}

\begin{figure}[t]
  \centering
 \includegraphics[width=0.99\columnwidth]{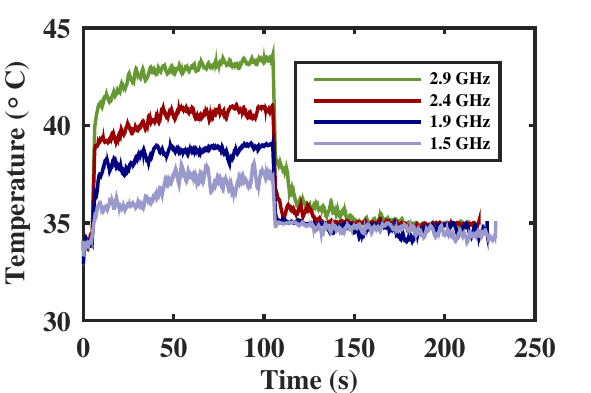}  
  \caption{\textbf{Effect of processor frequency on thermal behaviour:}
    Temperature profiles produced by running a CPU intensive application on a
    core at different processor frequencies for 100~s.}
  \label{fig:all-freq}
\end{figure}

\subsection{Example Temperature Trace}

To illustrate how computations affect the temperature of a core, we ran a CPU intensive application -- more specifically, one that does an RSA decryption continuously in a loop. We ran the application on core~3 of an octa-core processor (for setup details, refer to Section~\ref{sec:setup}).  Figure~\ref{fig:single-freq} shows the recorded temperature trace of core~3
during the execution of the application for 100~s (between the dotted red lines) on it and for about 50 s
thereafter when the core cools. We observe that 25~ms after the 
application begins execution, the temperature rises by 5$^\circ$C from
approximately 35 $^\circ$C to 40 $^\circ$C. Following this rapid rise, the temperature increases very slowly and saturates at 43$^\circ$C. As soon as
the application stops executing, the temperature falls rapidly to 38 $^\circ$C in about 25 ms and takes an additional 11~s to reach 35 $^\circ$C. 

The exponential nature of the temperature rise and fall is a result of
the system's thermal capacitance and resistance. In fact, the temperature fall
curve's characteristic allows the temperature changes caused by such an application's execution to be observed for sometime (in our example case, up to 11~s) after it has stopped.

\subsection{Factors Influencing a Core's \\ Temperature}
The major factors affecting the temperature at a particular core are
the fan speed, processor frequency and heat propagation from neighbouring cores. Since, in our
experiments, we do not control the server fan speed (see
Section~\ref{sec:setup}), we only discuss the effect of the processor frequency and
heat propagation on a specific core's temperature below.

\begin{figure*}[t]
  \centering
 \includegraphics[width=0.99\linewidth]{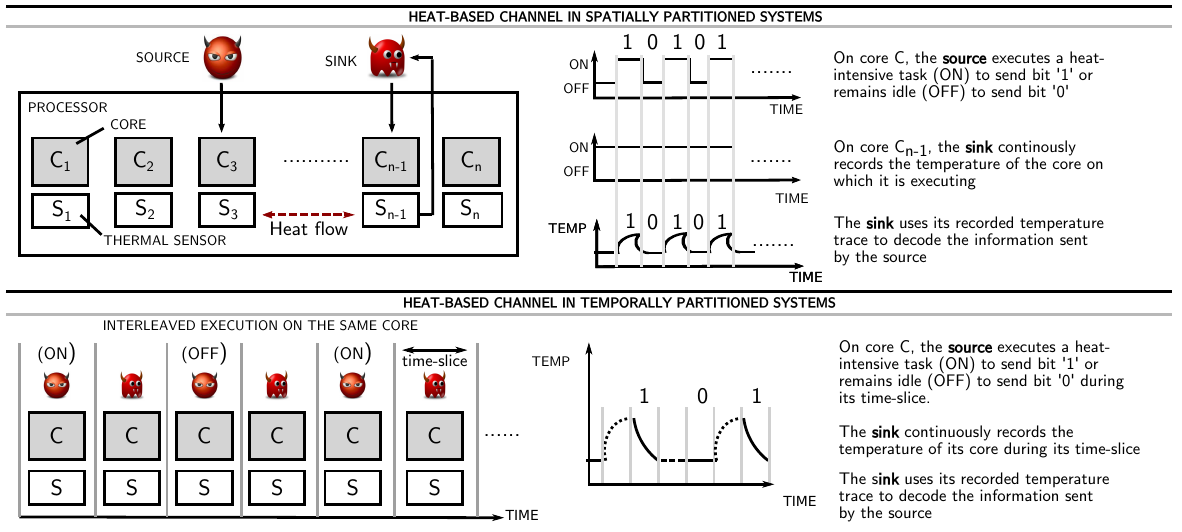}  
  \caption{\textbf{Covert communication using thermal channels:} We demonstrate that the temperature sensors on commodity multi-core platforms can be misused for covert communication by two colluding processes
    in spatially and temporally isolated systems. }
  \label{fig:goal}
\end{figure*}

\paragraph{CPU Frequency.} Most Intel processors are designed to run at a set of discrete frequencies for optimising power consumption. For example, in our setup (Figure~\ref{fig:setup}), the Xeon server can
run at frequencies between 1.2 GHz and 2.9~GHz. All cores within a single processor chip
run at the same frequency. Changes in frequency at a given core are reflected across
all the other cores. The actual frequency can be controlled either by the user or by the
kernel; for example, Ubuntu systems allow users to control this using the
\texttt{sysctl} interface~\cite{rubini1997sysctl}.

Figure~\ref{fig:all-freq} shows how the processor frequency affects temperature
when a CPU-intensive application runs for 100 s. We can observe that higher
frequencies result in more heat and higher saturation temperatures. 
This is because processor operation at a higher frequency results in a larger power density and therefore, more heat.

\paragraph{Heat Propagation From a Neighbour.} The heat resulting from computations on
one core will propagate to neighbouring cores. As a
result, the temperature at a certain core depends not only on that core's
workload (type of computation and schedule) but also those of its neighbours.

Figure~\ref{fig:hops} shows the effects of a CPU-intensive application executing on
a central core (core~3) of an octa-core processor for 100~s. We notice that the computation on core~3 affects the
temperature sensors of its neighbouring cores which remain idle all through.
Additionally, we observe that the saturation temperature of a neighbouring core
decreases with increasing distance from core~3. This effect is not symmetric as one would expect on either side of core~3. We suspect that this is due to an asymmetrically located processor hotspot.

\section{Exploiting Thermal Behaviour}
\label{sec:intro-channels}
In this section, we present the intuition underlying the construction of thermal side channels on multi-core systems which implement spatial or temporal resource partitioning for isolation.

\subsection{Isolation based on Spatial and \\ Temporal Partitioning}
Isolation techniques that rely on resource partitioning are becoming increasingly popular and there have been a number of proposals for using such partitioning to prevent covert channels~\cite{wang-acsac06,gundu-hasp14}. 
We consider two most common types of process isolation and partitioning techniques in our work: spatial and
temporal (Figure~\ref{fig:goal}). In spatially partitioned systems, processes are isolated by being assigned 
exclusive computation resources, i.e., no two processes share cores or memory. 
Such an approach prevents certain types of side-channels between processes that execute concurrently. For example, cache-partitioning prevents any information leakage that may occur based on state of the cache-lines in a processor (e.g., how many cache-lines are full). In such systems, processes do not share any processor temperature sensors because they do not use any common CPU resources.

In temporally partitioned systems, the processes share the same resources but
run in a time-multiplexed manner. For example, this technique is used by TEEs like Intel TXT in which only one of  two partitions (trusted or untrusted) are active at a time but have access to common cores and memory.  In systems that employ temporal partitioning,  processes that share one or more cores have access to the corresponding temperature sensor(s) during their execution time-slice.

Thermal side channels that leverage system thermal behaviour can be used to circumvent both these types of isolation techniques as we describe below.

\subsection{Constructing Thermal Channels}
Based on our discussion in Section~\ref{sec:therm-behavior}, we make two observations that can be used to construct thermal side channels. 
\begin{enumerate}
\item \textbf{Remnant Heat.} Since temperature variations that result from a computation can be observed even after it stops, these variations leak information regarding the computation to the process that follows it in the execution schedule especially if it is on the same core. This remnant heat can be exploited as a thermal side channel and may allow a process to exfiltrate sensitive information from its predecessor thereby violating temporal partitioning. Furthermore, it can also be used for communication between two colluding processes that time-share a core. Note that while it is possible to \emph{reset} most resources (e.g., CPU registers, caches) to prevent other side channels before switching between applications, the remnant heat from a computation (and hence, a thermal side-channel) is hard to eliminate.
\item \textbf{Heat Propagation to a Neighbouring Core.} The thermal conductivity
  of the processor results in heat propagation between cores, i.e, the heat that
  results from a computation not only affects its underlying core's temperature
  but also its neighbouring cores. As a result, this heat flow can be exploited as a thermal channel by an attacker to either make inferences about a potentially sensitive computation at a neighbouring core or by colluding processes to communicate covertly thereby violating spatial resource partitioning. Since it is hard to eliminate heat flows within processors, thermal side channels remain a viable threat even in spatially partitioned systems.
\end{enumerate}

There are several challenges involved in the construction of thermal side
channels. First, the nature of temperature changes makes it hard to control the effect that an application's execution will have on the temperature of its own core and its neighbours. Second, the limited resolution of the temperature sensors available on current x86 platforms prevents fine-grained temperature monitoring. Finally, fan-based cooling mechanisms affect the rate and extent of temperature variations. 

\begin{figure}[t]
  \centering
 \includegraphics[width=0.99\columnwidth]{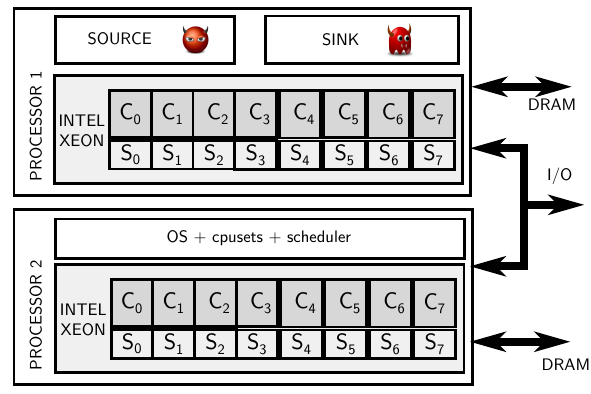}  
  \caption{\textbf{Our Experimental Setup:} Our framework consists of an Intel Xeon-based server platform running
    Suse Linux. We use \emph{cpusets} to achieve spatially or temporally isolated source and
    sink applications. We wrote a custom source application that uses RSA decryption operations to generate heat and a sink application that records its own core's temperature continuously.}
  \label{fig:setup}
\end{figure}

 \section{Covert Communication Using Thermal \\ Channels}
 \label{sec:covert}
 In this section, we present the feasibility and throughput of communication using thermal covert
channels in spatially and temporally partitioned multi-core systems. We begin by describing our experimental setup that implements such isolation mechanisms below. Throughout, we refer to the data sender as the \emph{source} and the recipient as the \emph{sink}.
 
\subsection{Experimental Setup}
\label{sec:setup}
Our setup is based on an Intel server containing two octa-core Xeon processor chips and running an Open SUSE installation (Figure~\ref{fig:setup}). We use \emph{cpusets} \cite{cpusets}  to implement spatial and temporal partitioning. Using this,
we restrict the OS to one of the processor chips (Processor 2) and isolate it from the rest of
the system. We achieve spatial partitioning by running the source and the sink on separate cores on the Processor 1 with minimal interference from the OS. To realise temporal partitioning, our system
incorporates a scheduler that controls the duration and cores on which the source and sink execute.

We wrote a custom application that performs an RSA
decryption (using PolarSSL~\cite{polarssl}) continuously in a loop and use it as the source application of the covert channel. Similar to popular thermal benchmarks (e.g., CStress~\cite{cstress}, mprime~\cite{mprime}), this application extensively uses
the CPU register file and ALU and can therefore, increase the temperature of its underlying core quickly.  

We rely on the server fan for cooling the cores. Our server allows the fan-speed
to be configured only through the BIOS and we set this to a maximum speed of
15000 rpm for our entire study. We chose this setting because it is the most likely
setting for servers which run computationally intensive tasks. Our server is currently in a
room whose ambient temperature is around 22$^{\circ}$C. We also implemented  a
custom sink application that records the temperature of the core on which it executes continuously. 

Our experimental framework is implemented using C and allows configuration of run-time
parameters like the processor frequency, set of applications to run, their
schedule and mapping to cores. Initialisation and tear down of the measurement framework
is performed using a set of Perl and Bash scripts. Our setup allows us to
achieve strong spatial and temporal isolation; this makes it an ideal platform
for our investigation of thermal side channels.

\subsection{Covert Communication in Spatially Partitioned Systems} 
\label{sec:exp-spatial}
This section addresses the construction of thermal channels in the scenario where the source and sink
applications run on dedicated cores and execute concurrently. The sink has
access only to its own core's temperature sensor and not that of the source as
described in the upper part of Figure~\ref{fig:goal}. To communicate covertly in such a scenario, the source exploits the heat propagation from its own core to the sink that runs on a neighbouring core. In this section, we demonstrate the feasibility of achieving this on a commodity multi-core platform and evaluate the throughput of such a communication channel. Below, we first present the encoding scheme that we use for data transmission and then describe the experiments that realise covert communication using the thermal channel.

\paragraph{Encoding and Decoding.} The source and sink use the \emph{ON-OFF Keying} for their communication. To send bit `1', the source application runs RSA decryption operations to
generate heat and to transmit bit `0', it remains idle. It is important that the source application runs long enough to affect the sink's
temperature sensor on a neighbouring core to send bit `1', i.e., it must
generate enough heat to raise the temperature of the sink's core 
above the ambient temperature. We denote the the minimum
duration for which the source application needs to execute to transmit a
`1' bit to the neighbouring cores as  \bitlen. The source
remains idle for the same duration to send a `0' bit. We assume that the source and sink agree on \bitlen{} and a fixed preamble to mark the start of the data apriori.

The sink that records the temperature of its own core continuously does the following to decode the data. It first searches the recorded temperature trace for a fixed preamble. We choose a preamble starting with bit `1' because it can clearly be identified by the sink. To detect the start of the preamble, the sink searches for a `1' bit by detecting the first rising edge, i.e., a temperature increment $\geq$2$^\circ$C given its ambient temperature. We use this threshold because the resolution of the sensors on the platform is $\pm$1$^{\circ}$C. It then tries to decode the bits following this to see if they match the preamble. The source repeats this until it recovers the preamble from the temperature trace.  It then decodes the remaining bits using a simple edge detection mechanism in which a rising edge indicates  bit `1',  a falling edge indicates bit `0' or  a no-change implies that the value is the same as the previous bit. 

\begin{figure}[t]
  \centering
 \includegraphics[width=0.99\columnwidth]{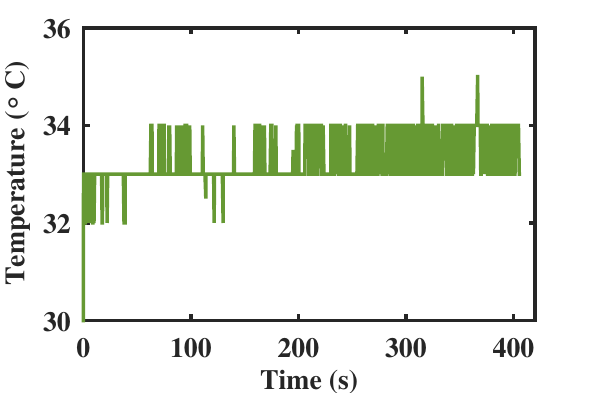}  
  \caption{\textbf{Temperature drift due to the sink's execution:} The temperature drift caused by the execution of the sink itself is
    very slow as shown here. This trace was recorded over 400~s by running the
    sink application on core~3 with all the other cores idle. Note that the
    resolution of the sensor is $\pm$1$^{\circ}$C.}
  \label{fig:observer_trace}
\end{figure}

\begin{figure}[t]
  \centering
 \includegraphics[width=0.99\columnwidth]{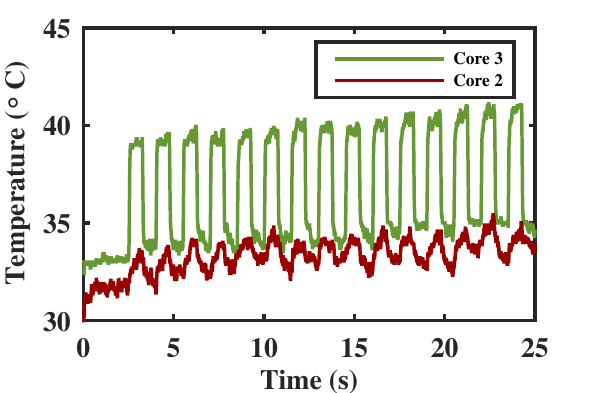}  
  \caption{\textbf{Thermal communication in spatially isolated systems:} Temperature traces recorded during the transmission of the 30 bits (15 ones and 15 zeroes) from the source (core~3) to the sink (core~2) using a \bitlen{} of 750~ms.}
  \label{fig:hop1_temp}
\end{figure}

\paragraph{Temperature Drift due to the Sink's Execution.} The sink relies on the source to affect its core's temperature sensor for communication. However, to do this, it is necessary to isolate any temperature drifts that may be caused by the sink's execution itself on its own temperature sensor. 

To understand these drifts better, we run the sink for a long time and observe its temperature. Figure~\ref{fig:observer_trace} shows the temperature trace of core~3 when the
sink is running on it for 400 s while the other cores are idle. The core's temperature
remains stable at around 33$^\circ$C for 200 s and later drifts slowly towards
34$^\circ$C. Therefore, we conclude that the temperature changes caused by the execution of the sink process itself is
negligible over a long duration of time~(\eg 200 s).

\paragraph{Calibration of \bitlen.} Before the actual transmission of data, we have to determine the optimal value of \bitlen{}, i.e., the duration for which the source executes or remains idle to send bit `1' and bit `0' respectively. Note that the actual value of \bitlen{} depends on the relative locations of the source and the sink. This is because the effect of the source's execution affects the cores farther away from it to a lesser extent (see Section~\ref{sec:therm-behavior}). For our first experiments, we fix the source to execute on a core~3 because it is a central core and the sink to execute on core~2. We later describe the effects of increasing the distance between the source and the sink on \bitlen{}. 

To estimate \bitlen, we first set it to a value between 50 ms and 1500 ms. We then attempt to send 100 data bits from the source on core~3 to the sink on core~2 and observe the resulting temperature traces on core~2. We do this by configuring the source application to be active and stay idle for \bitlen{} alternately. Our data bits consists of 50 alternating ones and zeros. We choose this data sequence because it is important to ensure that the chosen \bitlen{} consistently results in the desired temperature increment on core~2.
 
We were unable to decode data when \bitlen{} was smaller than 250 ms. We observe that data transmission using  \bitlen$\geq$ 500 ms results in about 10\% bit errors (Table~\ref{tab:error_hop}). Furthermore, we notice that the bit error rate does not improve much by increasing \bitlen{} from 500 ms to 1500 ms. Figure~\ref{fig:hop1_temp} shows the temperature traces of the two cores during the data exchange using a \bitlen=750~ms. The data shown here has been post-processed to remove noise using a smoothing function. We observe that the temperatures of core~3 and core~2 are well-correlated (correlation co-efficent $\simeq$ 0.6 \%, p-value = 0).

\begin{table}[t]
 \centering
 \small
 \begin{tabular}{ |  r  |  c  |  c |}
 \hline
  \multirow{2}{*}{\bitlen (ms) }& \multicolumn{2}{|c|}{Bit Error (\%)} \\ 
 \cline{2-3}
  & Core 2 & Core 1  \\
  & (1-hop) & (2-hop)  \\
   \hline
 250  & 18  & -- \\
 \hline
 500 &  14 &  -- \\
 \hline
 750 & 13  &  --  \\
 \hline
 1000 & 11  & 24 \\
 \hline
 1250 & 9  & 26 \\
 \hline
 1500 & 8  & 15 \\
 \hline
 \hline
 \end{tabular}
 \caption{\textbf{Calibration of \bitlen{} in spatially partitioned systems:} We
   send a block of 100 bits consisting of alternating ones and zeroes using
   different \bitlen{} values from the source (core~3) to the sink that runs at
   one and two hop distances from it. The processor frequency was set to 2.9~GHz and
   this table shows the resulting bit-error rates (`--' indicates that the data could not be decoded). We
   observe when \bitlen$\geq$ 500 ms, we can decode data with less than 15\% error
   at one hop but this does not improve much by increasing \bitlen{} to 1500~ms. We
   also notice that the required \bitlen{} increases with greater distance from
   the source. At a one hop distance, setting \bitlen=750ms and using
   Hamming(7,4) error correction code results in a channel throughput of up to
   \textbf{0.33bps}. 
 }
 \label{tab:error_hop}
 \end{table}

\paragraph{Error Rate.} To understand the nature of errors in thermal channels,
we send a pseudorandom sequence of a 1000 bits in 100-bit blocks. Each block
begins with a preamble to enable the sink to detect the start of data transmission. 

From our initial experiments, we observe that the temperature traces of core~2
and core~3 are well-correlated in time over a sequence of alternating ones
and zeros (Figure~\ref{fig:hop1_temp}). Therefore, we choose a preamble of five
alternating ones and zeroes (10 bits in total).  The source and
sink synchronise in 9 out of 10 tests and the average error rate is 13.22
\%~($\pm$~5.19) for a \bitlen{} value of 500 ms. On increasing \bitlen{} to 750 ms and 1000 ms, the source and sink synchronise over all 10 tests and the average error rate is  11.3\%~($\pm$~2.83) and 11\%~($\pm$~3.83) respectively.

\paragraph{Varying the Sink's Location.}
We repeat similar experiments by running the sink in core~1 and core~0 which are
two and three hops away from the core~3 to see how the error rate varies with
increasing distance from the source. At a two hop distance, we observe that for
a given \bitlen{}, the error rate is higher than in the case of the one hop
(Table~\ref{tab:error_hop}). At a three hop distance, we were unable to decode
data at 1500 ms. However, increasing the value of \bitlen{} sufficiently will
allow data transmission at a 3-hop distance. For example, Figure~\ref{fig:hop3}
shows an extreme case in which \bitlen{} is set to 200 s to transmit bit `1'.

The increased error rate and deterioration in the ability to decode data  is expected because heat resulting from computations at a given core affects the cores closer to it more than the cores farther away.  We repeated the experiments to estimate the error rate from the source (core~3) to a sink running on core~1  at a two hop distance. We observe that the source and sink synchronise successfully in 9 out of 10 tests. We can transmit data at the rate of 1 bit in 1.5 s (\bitlen{} = 1500 s) with an error rate of 18.33\%~($\pm$4.21).

\paragraph{Effect of Frequency on \bitlen.} To understand the effect of
processor frequency on \bitlen{}, we repeated our experiments for 1-hop
communication at a lower frequencies, namely, 2.4 GHz and 1.9 GHz. As shown in
Table~\ref{tab:error_freq}, for a given \bitlen{}, the error rate increases at
lower processor frequencies. When the processor frequency is set to 1.9~GHz, we
could not decode data even at 1500 ms. We note that using a larger value for
\bitlen{} would solve this problem and can be done using the same methodology we
used for our experiments. This deterioration in error rates and the ability to
decode data itself is expected because lower frequencies result in lesser heat
generation from a given computation and therefore, the rise in temperature may
not be significant enough to detect a bit `1' . 

We repeated the data transmission experiments when the processor frequency is set to 2.4 GHz. We transmit a pseudo-random sequence of 1000 bits in 100-bit blocks, each prepended with a preamble, from the source (core~3) to the sink (core~2) using \bitlen=1250~ms and 1500~ms. In both cases, the source and sink sychronise in all 10 tests. The observed error rates in both cases is similar, i.e., 14.9\% ($\pm$~3.9) and 15.9\% ($\pm$~6.08) for \bitlen=1250~ms and \bitlen=1500~ms respectively.

\paragraph{Throughput Estimation.} 
From the above discussion, we conclude that the throughput of thermal covert channels in spatially partitioned systems depends on number of factors including the time required to transmit one bit of information (\bitlen{}) and error rates; both these parameters in turn depend on the processor frequency and the distance between the colluding processes. 

At 1-hop distances, given a \bitlen{} of 750 ms, the throughput would be 1.33
bps in the ideal case without any errors. However, due to the 11\% errors that we observe in the experiments, actual communication would require error correction to be implemented.  When we analysed the nature the errors, we found that for every four bits, with a probability of over 0.9, there was one or no errors. 
Therefore, we could use a  Hamming~(7,4) error-correction code to correct for these errors and this would result in  
75\% overhead and hence, an effective throughput of 0.33~bps. When the frequency is changed to 2.4~GHz, the throughput is about  0.2 bps using a Hamming~(7,4) error correction code. A similar trend was observed on increasing the distance between the source and sink. 

Finally, we note that the actual throughput of such a thermal communication channel in particular systems will depend on the actual machine and its workloads. Our experiments were performed in conditions that minimise any noise from the OS,  other processor workloads (e.g., only the source and the sink are active), etc. and the above estimates are representative upper bounds on the throughput of thermal channels on such a platform with spatial partitioning. Any additional processes that run alongside the source and the sink are likely to cause more communication errors depending on their relative location with respect to the source and the sink because they would contribute to the temperature variations at the sink's core and hence, lower the throughput.

\begin{table}[t]
 \centering
 \small
 \begin{tabular}{ |  r  |  c  |  c  |}
 \hline
  \multirow{2}{*}{\bitlen (ms) }& \multicolumn{2}{|c|}{Bit Error (\%)} \\ 
 \cline{2-3}
  & 2.9 GHz & 2.4 GHz\\
   \hline
 250  & 18  &  -- \\
 \hline
 500 &  14 &  23 \\
 \hline
 750 & 13 & 24 \\
 \hline
 1000 & 11  & 23 \\
 \hline
 1250 & 9  & 14 \\
 \hline
 1500 & 8  & 14  \\
  \hline
\hline
\end{tabular}

\caption{\textbf{Effect of processor frequency on required \bitlen:} We send a
  block of 100 bits consisting of alternating ones and zeroes using different
  \bitlen{} values from the source (core~3) to the sink (core~2). The table
  shows the resulting bit-error rates at different processor frequencies  (`--'
  indicates that the data could not be decoded). We observe that when the
  processor runs at lower frequencies, \bitlen{} has to be increased to achieve lower bit-error rates.}
  % title of Table
 \label{tab:error_freq}
 \end{table}

 \subsection{Covert Communication in Temporally Partitioned Systems}
 \label{sec:exp-temporal}
Temporal partitioning schemes securely multiplex the same resources (e.g., cores, memory) between several applications. Systems that use this technique mitigate information leakage through side channels by clearing caches, registers, \etc while switching between processes. However, the thermal footprint of an application (the source in our case) remains intact for observation by the other application that executes after it on the same core (the sink in our case). This is a result of the thermal capacitance and resistance of processors and can be exploited to communicate covertly. We demonstrate this through our experiments below.

\paragraph{Scheduling, Encoding and Decoding Schemes.} In temporally partitioned systems, a scheduler determines the order in which different partitions execute on a core and therefore, we implement a scheduler (Figure~\ref{fig:setup}) that realises this functionality. Since the sink and source share the same core, they run in an interleaved manner and the sink has access to the temperature sensors only during its execution time-slice (\timeslice). Temperature variations over the course of the source's execution within one time-slice are not visible to the sink; instead the sink only has access to the final temperature after the source's execution time-slice. Therefore, in our implementation, the source sends a single bit per time-slice (using the temperature at the end of that time-slice) to the sink over a thermal covert channel. In other words, the bit length is equal to the the execution time-slice, \timeslice. We use the same ON-OFF keying technique as before in Section~\ref{sec:exp-spatial}.

\begin{figure}[t]
  \centering
 \includegraphics[width=0.99\columnwidth]{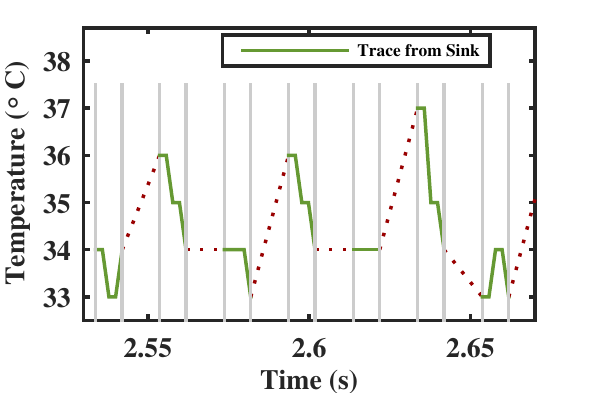}  
  \caption{\textbf{Thermal communication in temporally isolated systems:} Temperature traces recorded during the transmission of the 6 bits ( 3 ones and 3 zeroes) from the source  to the sink.  The source and the sink execute in alternate time-slots of 10~ms (marked in grey) on core~3. The thick lines are the actual temperature traces recorded by the sink and the dotted lines represent the temperature changes that occur as a result of the source's execution.}
  \label{fig:hop0_temp}
\end{figure}

\paragraph{Calibration of \bitlen.} We consider the scenario in which the sink and the source share a core and run in a round-robin fashion. The source heats up the processor (bit `1') or stays idle (bit `0')  to send one bit of information to the sink application that runs immediately after.

In order to understand how fast one can transmit bits over such a channel, we do the following. We vary \bitlen{} between 10 ms (which is the minimum value that our framework allows) and 30 ms and try to send an alternating sequence of 50 ones and 50 zeroes from the source to the sink. The source and sink run on core~3 in our experiments. Figure~\ref{fig:hop0_temp} shows an example temperature trace that the sink records during its execution. Note that the sink has access to the shared core's temperature sensor only during its own time-slice. We observed no errors in the data that the sink decodes even for \bitlen$\geq$10 ms and therefore, use this value for further experiments. We also repeated the experiment on the cores at the corners (core~0 and core~7) and noticed similar results.

\paragraph{Error Rate.} To understand the nature of errors in this channel, we
send a pseudorandom stream of 1000 bits in 100-bit blocks using \bitlen=10 ms.
We send 10 bits for synchronisation at the beginning of every block similar to
the experiments in Section~\ref{sec:exp-spatial}. The synchronisation using the preamble was successful in all cases and the data transmission resulted in error rates of  7.6\% ($\pm$ 1.9),  9.5\% ($\pm$ 4.86) and 7.1\% ($\pm$ 2.23) for experiments on cores~0, core~3 and and core~7 respectively.

 \begin{table}[t]
 \centering
 \small
 \begin{tabular}{ |  r  |  c  |  c  | c |}
 \hline
  \multirow{2}{*}{\bitlen (ms) }& \multicolumn{3}{|c|}{Bit Error (\%)} \\ 
 \cline{2-4}
  & 2.9 GHz & 2.4 GHz & 1.9 GHz\\
   \hline
 10  & 0  &  4 & 0 \\
 \hline
 15 &  0 &  1 & 1\\
 \hline
 20 & 0  & 0  & 1\\
 \hline
 25 & 0  & 0 & 0\\
 \hline
 30 & 0  & 0  & 0\\
  \hline
\hline
\end{tabular}
\caption{\textbf{Effect of processor frequency on required \bitlen:} We send a
  block of 100 bits consisting of alternating ones and zeroes using different
  \bitlen{} values from the source (core~3) to the sink that runs on the same
  core. The table shows the resulting bit-error rates at different processor
  frequencies. We observe that the error rates do not change much even at lower
  processor frequencies. Setting the frequency to 2.4 GHz, \bitlen{} to 10 ms
  and using Hamming(7,4) error correction code leads to the channel throughput
  of up to \textbf{12.5bps}.}
  % title of Table
 \label{tab:error_hop0_freq}
 \end{table}
 
 \paragraph{Effect of processor frequency on \bitlen.} We repeated our experiments at two lower processor frequencies (2.4 GHz and 1.9GHz) to understand how it may affect the required \bitlen{} for reliable communication. In the \bitlen{} calibration experiments, the error rates remain low despite the decrease in frequency (Table~\ref{tab:error_hop0_freq}). In the actual data transmission tests (of 1000 bits in 100 bit blocks) at 2.4 GHz, the source and the sink successfully synchronise in all 10 tests and the error rate is about 6.5\% ($\pm$3.58). At 1.9 GHz however, the synchronisation succeeds only 5 out of 10 times and the error rate is 9.5\% ($\pm$2.55). This indicates that a higher \bitlen{} value is required for more reliable communication at a processor frequency of 1.9 GHz.
 
 \paragraph{Throughput Estimation.} The throughput of the thermal channel in temporally partitioned systems depends on the execution time-slice per application, their schedule and the time required to send one bit of information~(\bitlen). If the execution time-slice (\timeslice) $\le$ \bitlen{}, then communication becomes difficult because the source cannot generate enough heat to transmit `1' bits. However, if \timeslice{} $\geq$ \bitlen{}, then the source can choose to execute for long enough to cause a temperature change  that the sink notices. We note the typical Linux systems have a time-slice of about 100 ms which is 10 times bigger than the one we need for implementing thermal covert channels.
 
 When \bitlen{} (also, equal to \timeslice) is 10 ms, we would expect the throughput of the thermal channel would be 50 bps. However, since the communication is error prone and results in up to 10\% error, the encoding scheme would have to incorporate error correction codes. On analysing the nature of the errors during the transmission of a 1000 bits, we see that with a probability of over 0.9, there is 1 or no errors in every four data bits. Therefore, we can use a Hamming(7,4) code to overcome these errors and this results in an effective throughput is about 12.5 bps.  This throughput is independent of which particular core the source and sink share (core~0/3/7). A \bitlen{} of 10~ms results in low error rates even at a processor frequency of 2.4 GHz and hence, the throughput is roughly 12.5 bps.
 
 We note that the actual throughput of  thermal channel in a particular system would depend on its hardware and its workloads. In our experiments, we minimise noise from running the source and sink in isolation from the rest of the system (OS, other workloads, etc.) and therefore, the results we present are representative of upper bounds on the achievable throughput on our platform implementing temporal partitioning. This is because any additional workloads that run along with the sink and source processes would affect the temperature traces that the sink records and hence, lower the throughput.
 
 \section{Thermal Channels for Unauthorised Profiling}
 \label{sec:profiling}
 In this section, we present a preliminary study of how thermal side channels enable unauthorised thermal profiling of processes even in systems that implement strong isolation mechanisms like spatial resource partitioning. 
 
Since the heat generated from an application (\emph{victim}) can be observed from its neighbours, this may leak information regarding the nature of its computation to processes at other cores. More specifically, if the attacker has reference thermal traces for   victim applications, he can recognise if and when such an application executes on a neighbouring partition. 
For example, identifying that a sensitive or potentially vulnerable application is running on a neighbouring core may aid an attacker in preparing or launching an attack. Application identification based on temperature traces has not been addressed previously in literature and below we present a first study that tries to understand its effectiveness as an attack vector. 

\paragraph{Goal and Intuition of the Attack.} We assume that an attacker has
access to a reference thermal trace of the victim application (say RSA
decryption); such a trace can be obtained by the attacker if he has access to a
similar platform as the one he is attacking. The attacker's goal is to verify if
the temperature trace of his core is a result of the victim application's
execution on a neighbouring core. Note that the attacker does not have access to
the temperature trace of its neighbouring core(s) but only to that of his own core. For
simplicity, we assume that only the attacker and the victim are active during
the attack and that they are collocated on adjacent cores. The attacker
continuously monitors his own core's temperature and then, correlates it with a
reference trace of the victim application. A strong correlation indicates that
the attacker's temperature trace was a result of the execution of the victim
application with high probability.

\paragraph{Experiments and Analysis.} We chose a set of five CPU-intensive applications including RSA decryption and four applications from a benchmark suite, MiBench~\cite{guthaus-wwc01} (ADPCM, Quick Sort, BitCount, BasicMath) and use them as the universal set of applications that a victim core (core~3) executes. We intentionally chose similar applications all of which stress the ALU and register file region of the core because this is harder than distinguishing between applications that are very distinct, for example, ones which stress the caches as opposed to the register file. 

 \begin{figure}[t]
  \centering
 \includegraphics[width=0.99\columnwidth]{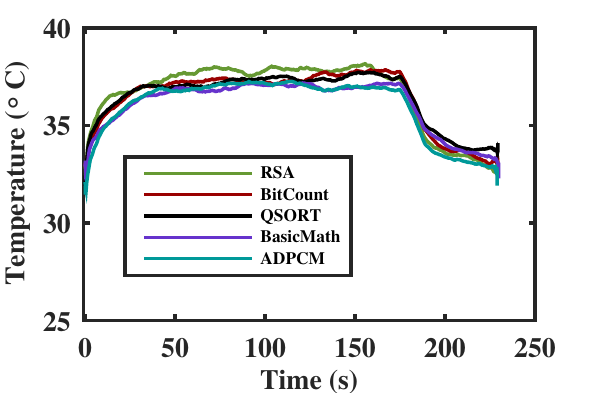}  
  \caption{\textbf{Example Temperature traces of different applications:} We ran a set
    of five CPU intensive applications (RSA decryption, BitCount, QSort,
    BasicMath, ADPCM) from a popular MiBench benchmark
    suite~\cite{guthaus-wwc01} for 200~s one each at a time and recorded the resulting temperature traces on a neighbouring core.}
  \label{fig:all-app}
\end{figure}

To understand the feasibility of identifying these applications, we ran each of them for 200 ms on core~3 of our setup (Section~\ref{sec:setup}) and collected the temperature traces of a neighbouring core (core~2). We repeated this five times for each of the applications and Figure~\ref{fig:all-app} shows one such trace for each of them. We observe that the saturation temperature for each of the traces is different. 

We use a simple correlation as a metric to measure similarity/differences
between pairs of applications. On correlating the traces from the RSA
application, we observe that the correlation is higher than 85\% in seven out of
ten (since there are 5 runs, we have 10 pairs to correlate) occasions. Using
this same correlation threshold of 85\% also results in 28\% false positives
when the RSA application is correlated with the others from the benchmark suite.
In general, traces belonging to the same application have high correlation
values ($\geq$80\%). However, traces belonging to different applications also
show high correlation because they are all CPU intensive and stress similar
parts of the CPU ($\geq$75\%). Therefore, we conclude that using a simple
correlation metric would only allow distinguishing applications that behave very
differently and more sensitive metrics (such as thermal models~\cite{rai-date15}
or machine-learning based classifiers) are required for better accuracy in other cases.

Finally, more fine-grained data exfiltration (e.g., deducing AES or RSA keys) on commodity x86-systems using the thermal side-channel is an open, unexplored problem. A key challenge is the limited resolution of the temperature sensors which is $\pm$1$^{\circ}$C and the rate at which the sensors are refreshed (currently, once every 2 ms).

\section{Discussion}
\label{sec:discussion}

In this section, we present possible countermeasures against thermal covert-channels and discuss their limitations. We also discuss  a security-application that can leverage thermal side channels.

\paragraph{Countermeasures.} Since we leveraged the temperature information exposed to software to construct thermal side channels, a 
natural solution to this problem would be to restrict access to temperature sensors on the system. However, such information cannot always remain hidden.  For example, in cloud systems, it is important for a guest operating system to track temperature information not only to schedule intelligently but also to understand if any of its user processes are misbehaving. Furthermore, centralised control and monitoring of thermal states does not scale well with  the advent of many-core processors~\cite{tilera, xeonphi} that contain hundreds of cores and host a large number of autonomous processes on separate cores, In fact, research prototypes like Intel's SCC platform~\cite{intel-scc} already allow subsets of cores to administer their frequency and voltage independently for power-efficiency; temperature information is a vital input to this decision process. \emph{Therefore, there is a clear tension between securing platforms and improving their energy-efficiency by exposing thermal data to the software on them.}

Even if one restricts access to the temperature sensors, related information such as clock skew, fan speed and even frequency in systems that allow dynamic frequency scaling still leak information about the system's thermal state. Since all these parameters are usually common across cores or subsets of cores within a processor chip, they can still provide a signalling mechanism. Finally, while it may be possible to separate processes in time and spatially to limit the effectiveness of thermal channels, such resource allocation strategies are wasteful and result in low resource utilisation.

\paragraph{Thermal Fingerprinting For Security.} Although so far we discussed only how thermal behaviour of systems can be exploited by attackers, the same properties could be used for achieving better security. Since temperature changes resulting from computations are difficult to avoid, we hypothesise that thermal profiling techniques can also be used  to detect any anomalous behaviour in the execution of an application. More specifically, it is highly likely that run-time compromise of an application results in a temperature trace that does not match its original \emph{thermal fingerprint}.  It has been shown to be possible to extract thermal models 
by monitoring the application under controlled conditions~\cite{rai-cases12, rai-date15}.  By comparing the actual execution trace to the expected trace, it may be possible to detect run-time compromise of software applications but this needs further exploration.

\section{Related Work}
\label{sec:related-work}

We review previous work on temperature-related channels and other side-channels
in general on x86-systems. We also provide some examples of existing literature on optimizing
computing systems for thermal efficiency because it highlights the advantages of
exposing thermal data as opposed to the other work that misuses this data to
undermine security.

\paragraph{Thermal Channels and Attacks.}
There is no previous work that
demonstrates the feasibility of thermal side channels and their use for covert communication on commodity multi-core systems as we do in
this paper. Previously, two works discussed and one implemented thermal covert channels on FPGA boards ~\cite{marsh-ches07,iakymchuk-recosoc11,brouchier-iacr09}. There have also been attempts to transmit data between two processes by changing fan-speed~\cite{brouchier-sp09}. The ability to remotely monitor a system's clock-skew (influenced by the changes in the system temperature) has also been exploited in the past  for exposing anonymous servers ~\cite{murdoch-ccs06,zander-usenix08} and covert communication with a remote entity~\cite{zander-commletters13}. We note that although some of these works~\cite{brouchier-sp09,zander-commletters13} use the term thermal channel, none of them use the thermal information available on modern systems to covertly communicate between processes on the same host as we do in this paper. 

More recently, it has
been shown that it is possible to use temperature variations to induce processor
faults~\cite{rahimi-date12} which can in turn be also be used to extract
sensitive information like RSA keys~\cite{hutter-iacr14}. Thermal information can also be used for coarse-grained data-exfiltration. For example, since temperature directly reflects the intensity of computation, it can be used to estimate the load or resource utilisation of a machine. This was illustrated by Liu et al. who computed the resource utilisation of servers in Amazon's EC2 although in a completely different context using the temperature data that is exposed to virtual machines~\cite{liu-dasc11}. 

\paragraph{Temperature-based Denial-of-Service Attacks.} Previous research has
identified other security risks that arise from hardware and software thermal
management techniques on modern systems. For example, malicious processes may
cause a denial-of-service by slowing down the processor~\cite{hasan-hpca05} or
permanently damaging hardware by causing thermal hotspots~\cite{dadvar-stm05}. Such processes could also exploit the fact that certain architecture components (e.g. instruction cache) are ignored by thermal optimisation approaches on modern processors ~\cite{kong-dsc10}.

\paragraph{Covert channels on x86-systems.}
Originally defined by Lampson as part of the confinement problem~\cite{lampson-commacm73}, today,
several side channels have been identified and explored in the context of x86
systems. One of the most studied side-channels on x86-systems is the cache-based channel
which can be used for both covert data exchange~\cite{xu-ccsw11} and data
exfiltration~\cite{zhang-ccs14}.  Covert channels based on bus contention have also been shown to
be feasible~\cite{wu-usenix12}. Network stacks~\cite{mileva-jcs14}, file-system
objects~\cite{lampson-commacm73}, input devices~\cite{shah-usenix06} and virtual
memory~\cite{suzaki-eurosys11} have all been shown to support covert-channels.

A typical way to prevent the above side channels is to partition them and is
used for example to mitigate cache-based channels~\cite{wang-acsac06} and
bus-based channels~\cite{gundu-hasp14} (by reserving bus bandwidth in this
case). However, such partitioning techniques will not eliminate
temperature-based channels completely as we have shown in this paper.

\paragraph{Thermal Monitoring of Processors.}
Temperature management of computing systems has gained significance over the last
few years given the trend of increasing on-chip temperatures of modern processors. This has
resulted in efforts to design and implement better thermal management techniques
for processors including optimization of sensor placement
(e.g.,~\cite{nowroz-dac10,memik2008optimizing,mukherjee2006systematic}), improving algorithms for dynamic temperature
management (e.g.,~\cite{yeo-dac08}) and cooling
techniques~\cite{chu-dmr04}. There are also ongoing efforts to develop
frameworks to thermally profile applications~\cite{rai-cases12}, temperature-aware
schedulers~\cite{coskun-vlsi08,choi-islped07} and micro-architectures~\cite{skadron-sigarch03,lim-isqed02}. 

Thermal profiling can further be used to detect compromised process in embedded
systems~\cite{wolf-hal06} and design schedulers such that they do not leak
information through thermal fingerprints of applications~\cite{bao-trusted14}.

\section{Conclusion}
\label{sec:conclusion}
In this paper, we demonstrated the feasibility and potential of thermal side
channels on commodity multi-core systems.  We showed that such channels can be
built by exploiting the thermal behaviour of current platforms and can be used
to circumvent strong isolation guarantees provided by temporal and spatial
partitioning techniques. Our experiments indicate that it is possible to use
them for covert communication between processes and achieve a throughput of up
to 12.5 bps. We also demonstrated that thermal channels can be exploited to
profile applications running on a neighbouring core. 

Attacks based on thermal channels are further facilitated by the increasing
trend towards exposing system temperature information to users, allowing them to
make thermal management decisions for efficient operation. Our work therefore,
not only points out a serious limitation in the isolation guarantees that resource
partitioning techniques can provide but also highlights the tension between
designing systems to support user-centric thermal management for efficiency and
security.

\footnotesize \bibliographystyle{acm}
\bibliography{myref}
\end{document}